\documentclass[manuscript]{acmart}
\AtBeginDocument{%
  }

\usepackage{subcaption}
\usepackage{enumitem}
\usepackage[x11names]{xcolor}
\usepackage{quoting}
\usepackage{mdframed}
\colorlet{shadecolor}{LavenderBlush2}
\usepackage{lipsum}
\newenvironment{shadedquotation}
  {\begin{mdframed}[backgroundcolor=LavenderBlush2, leftmargin=0pt, rightmargin=0pt, innertopmargin=5pt, innerbottommargin=5pt]
   \quoting[leftmargin=0pt, vskip=0pt]}
  {\endquoting\end{mdframed}}
\usepackage[normalem]{ulem}
\begin{document}

\title{%
REFLECTing SPERET: Measuring and Promoting Ethics and Privacy Reflexivity in Eye-Tracking Research}

\author{Susanne Hindennach}
\authornote{Both authors contributed equally to this research.}
\email{susanne.hindennach@vis.uni-stuttgart.de}
\orcid{1234-5678-9012}
\author{Mayar Elfares}
\authornotemark[1]
\email{mayar.elfares@vis.uni-stuttgart.de}
\affiliation{%
  \institution{University of Stuttgart}
  \country{Germany}
}
\author{C\'{e}line Gressel}
\affiliation{\institution{University of T\"ubingen}
  \country{Germany}}
\email{celine.gressel@izew.uni-tuebingen.de}

\author{Andreas Bulling}
\affiliation{%
  \institution{University of Stuttgart}
    \country{Germany}}
  
\email{andreas.bulling@vis.uni-stuttgart.de}

\renewcommand{\shortauthors}{Hindennach et al.}

\begin{abstract}
The proliferation of eye tracking in high‑stakes domains -- such as healthcare, marketing and surveillance -- underscores the need for researchers to be ethically aware when employing this technology.
Although privacy and ethical guidelines have emerged in recent years, empirical research on how scholars reflect on their own work remains scarce.
To address this gap, we present two complementary instruments developed with input from more than 70 researchers: REFLECT, a qualitative questionnaire, and SPERET (Latin for "hope"), a quantitative psychometric scale that measures privacy and ethics reflexivity in eye tracking.
Our findings reveal a research community that is concerned about user privacy, cognisant of methodological constraints, such as sample bias, and that possesses a nuanced sense of ethical responsibility evolving with project maturity.
Together, these tools and our analyses offer a systematic examination and a hopeful outlook on reflexivity in eye‑tracking research, promoting more privacy and ethics‑conscious practice. 
\end{abstract}

\ccsdesc[500]{Security and privacy~Human and societal aspects of security and privacy}
\ccsdesc[300]{Human-centered computing~Ubiquitous and mobile computing systems and tools}
\ccsdesc[300]{Human-centered computing~Empirical studies in collaborative and social computing}
\keywords{Eye Tracking, Privacy, Ethics, Research, Reflexivity}

\received{10 November 2025}

\begin{teaserfigure}
    \centering
    \includegraphics[width=0.7\linewidth]{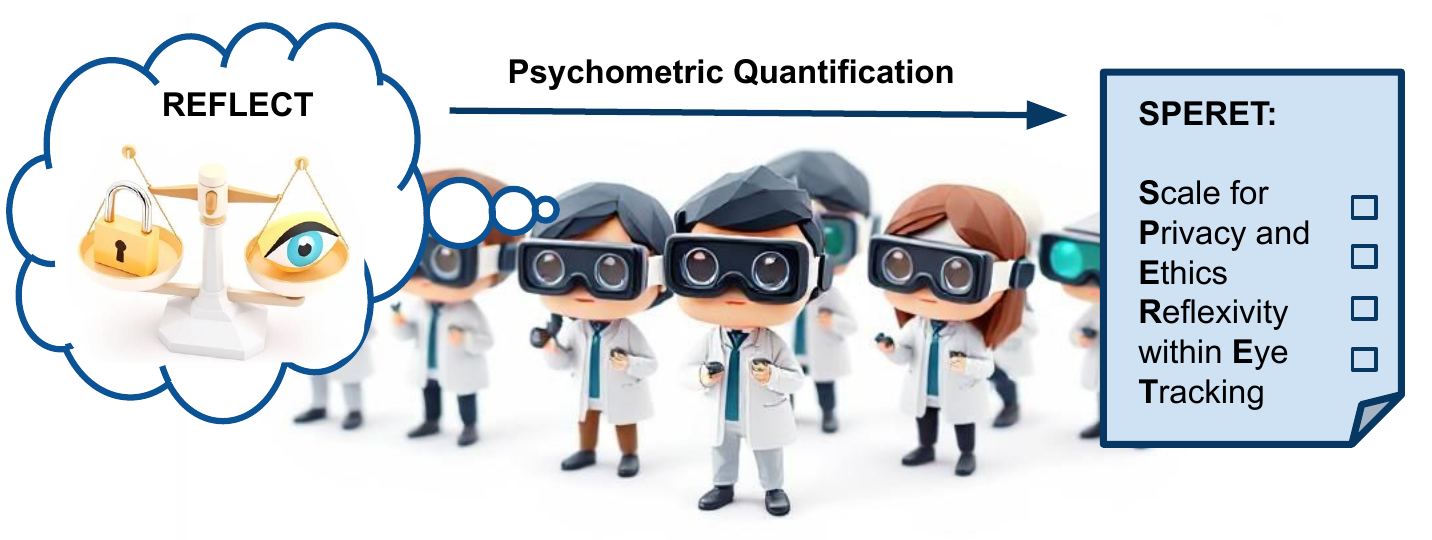}
    \Description[An overview of our approach.]{The image shows avatars with white lab coats and eye trackers that represent eye-tracking researchers. On the left side there is a thought bubble that represents the REFLECT questionnaire. Inside the thought bubble there is a scale representing ethics with a lock to represent privacy and an eye to represent eye tracking. On the right hand side there is a symbol for the SPERET scale. It is a sheet symbol with \textbf{S}cale to measure \textbf{P}rivacy and \textbf{E}thics \textbf{R}eflexivity within \textbf{E}ye \textbf{T}racking written on it, and checkboxes indicating that this is a item scale.} 
    \caption{%
    Using a novel REFLECT questionnaire, we asked eye-tracking researchers to reflect on the privacy and ethics aspects of their research. Based on an analysis of their reflections we further propose SPERET -- a psychometric \underline{S}cale to measure \underline{P}rivacy and \underline{E}thics \underline{R}eflexivity within \underline{E}ye \underline{T}racking.}
\end{teaserfigure}

\maketitle

\section{Introduction}
The falling cost and growing portability of eye-tracking hardware are shifting it from a specialised instrument used in psychology and human‑computer interaction labs to a pervasive technology employed across domains, such as marketing analytics, medical and psychiatric diagnosis, intelligent tutoring systems or next‑generation user interfaces \cite{bulling2016pervasive}.
The accompanying emerging social pervasion creates an urgent imperative to develop and deploy eye tracking responsibly \cite{liebling14privacy, gressel2023privacy}.
In this regard eye-tracking researchers so far mainly focus on privacy as it 
pertains to an individual’s right to control the collection, use, storage, and dissemination of their gaze data \cite{liebling14privacy, bozkir2023eye, gressel2023privacy}.
While privacy represents an important element of responsible research, broader ethical considerations also include issues of justice and fairness in participant sampling, as well as the anticipation and positively influencing potential consequences of the use of eye tracking technologies \cite{lavoie2025conducting, kaufmannEthicalChallengesResearching2021, hooge2025fundamentals, nystrom2025fundamentals, niehorster2025fundamentals}.

Numerous ethical guidelines have been proposed for conducting human‑subject research \cite{gostin1991ethical, rice2008historical, shamoo2021ethics} and for handling biometric data \cite{north2020biometric, sutrop2010ethical, alterman2003piece}.
These guidelines typically concentrate on procedural ethics -- such as informed consent and anonymisation -- while offering little empirical insight into the informal ethical reasoning, socio‑technical assumptions, and perceived responsibilities of the researchers who actually develop the methods and systems.
Researchers determine the project scope, select the technologies, and often decide whether and how to involve ethicists \cite{spindlerSituatedMultimodalityIntegrating2025}.
Even when ethicists are included, their input usually arrives at later stages of the project \cite{spindlerSituatedMultimodalityIntegrating2025}. Consequently, many ethical and privacy decisions are made by eye-tracking researchers who lack formal training in ethics or privacy, relying instead on unconscious assumptions and informal reasoning \cite{bieler2021distributing}.

This is despite the fact that ethical research demands that unconscious assumptions be continuously interrogated and aligned with established ethical theories \cite{vonungerEthicalReflexivityResearch2021}.
This practice -- stepping back from one's own work, questioning underlying premises, and conceptualising the factors that shape research unconsciously or contextually -- is known as \textit{reflexivity} \cite{bielerDistributingReflexivityColaborative2021, attiaBecomingReflexiveResearcher2017, vonungerEthicalReflexivityResearch2021}.
Beyond fostering ethically sound research, reflexivity also provides a methodological basis for empirically analysing researchers' perceptions of privacy and ethical issues.
This is essential because it offers a systematic means of assessing how researchers incorporate ethical and privacy considerations into their practice \cite{berghaeuserResearchersPracticePerception2025}.

The goal of this work is to foster and examine reflexivity in eye tracking research in order to promote ethical research practice.
To achieve this, we developed two complementary tools:

\begin{enumerate}
    \item \textbf{REFLECT} -- a qualitative questionnaire, and
    \item \textbf{SPERET} -- a psychometric scale 
\end{enumerate}

The REFLECT questionnaire elicits reflection via open questions about privacy and ethical aspects of research projects, while the SPERET scale offers a standardised means to assess researchers' reflexivity by asking for their agreement with validated items derived from the REFLECT insights.
A core part of development was the active involvement of eye-tracking researchers ($N > 70$) through offline workshops and online questionnaires, representing the perspectives of a diverse group of domain experts. 
This contrasts prior scale development \cite{farzandOutofDevicePrivacyUnveiled2024, hasan2023psychometric} in which items were generated through discussions among a small group of researchers or within a single research team (typically fewer than three researchers).
Complementarily, while REFLECT serves as an in-depth individual assessment that can help small teams to articulate their reflexivity, the standardised and quantifiable assessment of SPERET can enable broad assessment across larger research groups.
Together, they can identify gaps in ethical reflexivity, guide training or point out where an external ethical expertise is needed and thus open a connection point for interventions to enhance responsible research conduct.

We argue that our tools should be presented as developmental self‑assessment instruments intended to provoke reflection, dialogue and ethical growth, rather than as evaluative devices that judge moral integrity or determine eligibility for participation, funding or promotion.
Framed in this way, they retain a formative purpose, encouraging researchers to engage critically and honestly with their assumptions, values and decision‑making processes.
When incorporated into facilitated discussions or interdisciplinary ethics training, the tools become catalysts for collective learning, helping teams to uncover blind spots and negotiate divergent perspectives.
By contrast, using them as ranking or gate‑keeping mechanisms would undermine their reflective potential and encourage performative rather than genuine engagement.
Consequently, preserving a developmental framing is essential for the instruments to foster authentic ethical reflexivity and to contribute meaningfully to a responsible research culture.
\label{sec:intro}

\section{Related Work}
\subsection{Raising and Addressing Issues of Privacy and Ethics in Eye Tracking Research}
\label{sec_privacyrelatedwork}

Eye-tracking is increasingly being integrated into critical real-world applications, including driver monitoring in the automotive sector~\cite{braunagel2017dar,braunagel2017das}, diagnostic support in medicine~\cite{asfaw2018data, vidal2012wearable}, and assistive technologies~\cite{hwang2014eye}. Due to the richness of information encoded in human eye movements~\cite{krogerWhatDoesYour2020, bulling10_pcm}, across these domains, ethical, social, legal, and privacy considerations are paramount \cite{gressel2023privacy}. 
This concern arises from the limited voluntary control individuals have over their eye movements, which can unintentionally reveal sensitive personal, cognitive, or health-related information \cite{krogerWhatDoesYour2020}.

\begin{figure}
\centering
\begin{minipage}{0.5\textwidth}
  \centering
  \includegraphics[width=0.7\linewidth]{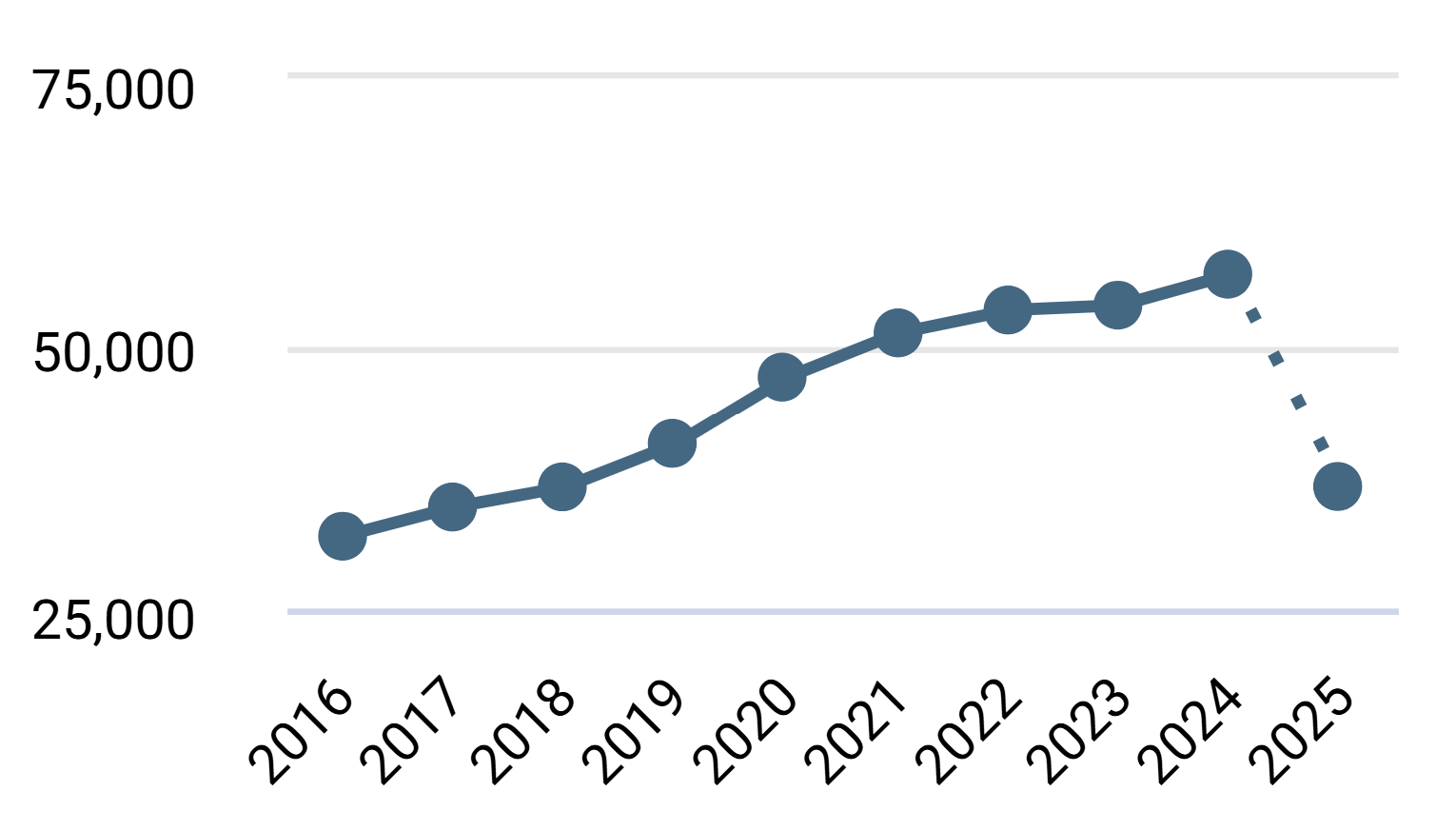}
  \label{fig:sub1}
\end{minipage}%
\begin{minipage}{0.5\textwidth}
  \centering
  \includegraphics[width=0.7\linewidth]{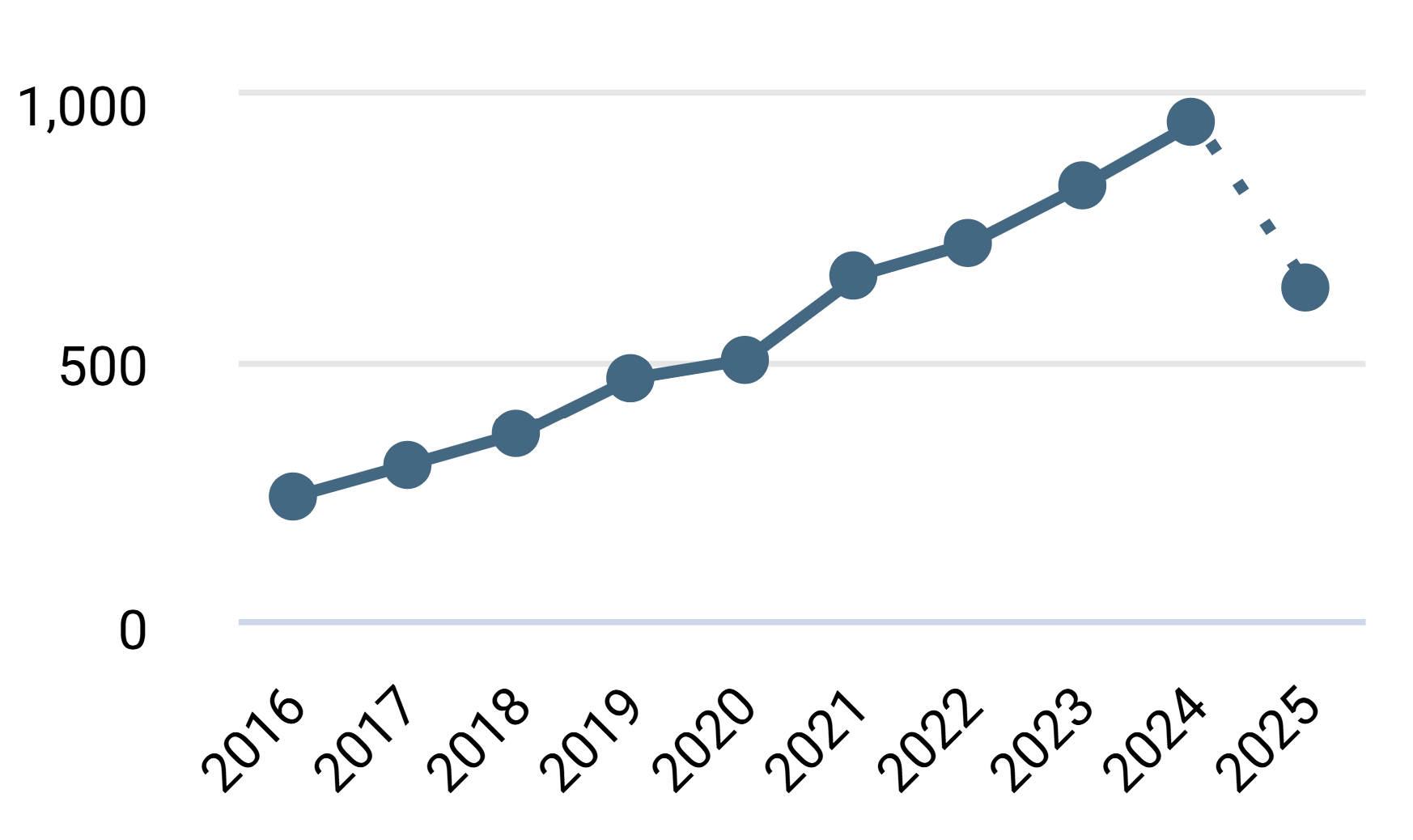}
  \label{fig:sub2}
\end{minipage}
\Description[Volume of publications in eye tracking overall, and publications that address ethics and privacy]{The figure shows to line charts for the publication volume over the years 2016-2025. The left line charts shows the overall volume. The line starts at ca. 30,000 in the year 2016 increases up to ca. 55,000 for the year 2024. The right line chart shows the volume of publications that address privacy and ethics aspects. It starts at 250 in 2016 and increases to 1000 in 2024.}
\caption{There exists a marked imbalance between the overall volume of publications in the field of eye tracking (Left) and the comparatively small number of studies that address the privacy and ethical dimensions of this technology (Right). The infographic is generated by dimensions.ai showing data from the past 10 years.}
\label{fig:stats}
\end{figure}

The proportion of eye-tracking studies that explicitly address or incorporate considerations of privacy and ethics remains relatively small in comparison to the total body of eye-tracking research, as shown in Figure \ref{fig:stats}.
Nonetheless, a growing body of research has emerged to address the privacy concerns inherent in eye-tracking technologies \cite{liebling14privacy, bozkir2023eye, gressel2023privacy, alsakarInvestigatingPrivacyPerceptions2023, steilPrivacyAwareEyeTracking}.
These studies mainly focus on gaze data collection and processing, including works on 
privacy-preserving gaze data sharing \cite{david2021privacy, elfares2023federated}, obfuscation techniques \cite{steil2019privacy, liu2019differential}, adversarial attacks \cite{hagestedt2020adversarial, sonnichsen2025attentionleak}, private quality verification \cite{elfares2025qualiteye}, and other formal or cryptographic solutions \cite{li2021kalvarepsilonido, elfares2024privateyes}.

Beyond technical solutions, several studies have explored the broader ethical, legal, and social implications of eye-tracking technologies 
\cite{gressel2023privacy, larsen2020ethical, kaufmannEthicalChallengesResearching2021, alsakarInvestigatingPrivacyPerceptions2023} %
focusing on research ethics trade-offs in eye tracking such as by-passers consent or data economy \cite{kaufmannEthicalChallengesResearching2021}.

\subsection{Reflexivity and Ethical Awareness Among Eye-Tracking Researchers}
Despite the growing recognition of privacy and ethical challenges in eye-tracking research, there remains little systematic understanding of how researchers themselves conceptualize, interpret, and address these issues in their own work.

Reflexivity as a necessary condition for producing socially robust, acceptable, and accountable knowledge, has been put forward in science and technology studies (STS) \cite{nowotnyDemocratisingExpertiseSocially2003}, critical HCI \cite{Bardzell2011}, feminist epistemology \cite{harawaySituatedKnowledgesScience1988}, Responsible Research and Innovation (RRI) ~\cite{IndicatorsPromotingMonitoring2015}, and other modes of interdisciplinary research. 
Accordingly, reflexivity constitutes an ethical and epistemological practice that enables researchers to situate their assumptions and values within knowledge production, thereby fostering anticipation of consequences and strengthening scientific rigor \cite{Bardzell2011, harawaySituatedKnowledgesScience1988}.

There has been an increase in anchoring reflexivity in research communities in the form of ethics or impact statements ~\cite{LiuEthicsStatements}. The main eye tracking conference Eye Tracking Research and Applications (ETRA) also introduced a dedicated \textit{Privacy and Ethics Statement} in 2024 (see  \autoref{ETRA_EthicsStatement} for the call). The initiative aims to raise researchers’ awareness, encourage reflection, and promote explicit discussion of the privacy and ethical implications associated with eye-tracking research and applications, nonetheless, its practical impact on researchers’ ethical reasoning remains unstudied.

\subsection{Privacy and Ethical Concern Measures and Scales}
Assessing how eye-tracking researchers engage with reflexivity requires dedicated instruments to measure ethical and privacy awareness, yet no such researcher-focused scale currently exists.
In psychology, HCI, and information systems research, several instruments have been developed to quantify privacy concerns among users. Prominent examples include the \textit{Out-of-Device Privacy Scale (ODPS)}~\cite{farzandOutofDevicePrivacyUnveiled2024}, the \textit{Value of Other People’s Privacy (VOPP)} \cite{hasan2023psychometric}, the \textit{Internet Users’ Information Privacy Concerns (IUIPC)} scale~\cite{malhotra2004iuipc}, and the \textit{Privacy Attitude Questionnaire (PAQ)}~\cite{ChignellPAQ}. 
These scales address user-oriented domains such as physical, online, or general data privacy, but no instrument currently assesses how eye-tracking researchers reflect on privacy and ethics within their own research practice.

While existing privacy scales are designed for users rather than researchers, prior efforts have examined researchers’ ethical perceptions across disciplines \cite{berghaeuserResearchersPracticePerception2025}. The study indicates that researchers generally possess a strong capacity to recognize ethical issues and regard them as integral to good scientific practice. However, such large-scale surveys lack the specificity required to address the unique privacy and ethical dimensions inherent to eye-tracking research.

Collectively, these studies and initiatives highlight the critical importance of integrating privacy and ethical considerations throughout the development and application of eye-tracking technologies. Sustained scholarly engagement and interdisciplinary dialogue remain essential to ensure responsible innovation. In this context, researchers themselves constitute the central pillars of ethical reflection and practice, as their awareness and decisions fundamentally shape how eye-tracking technologies are conceived, implemented, and applied in research and real-world contexts.\label{sec:related-work}

\section{Methodology}
\begin{figure}
    \centering
    \includegraphics[width=1\linewidth]{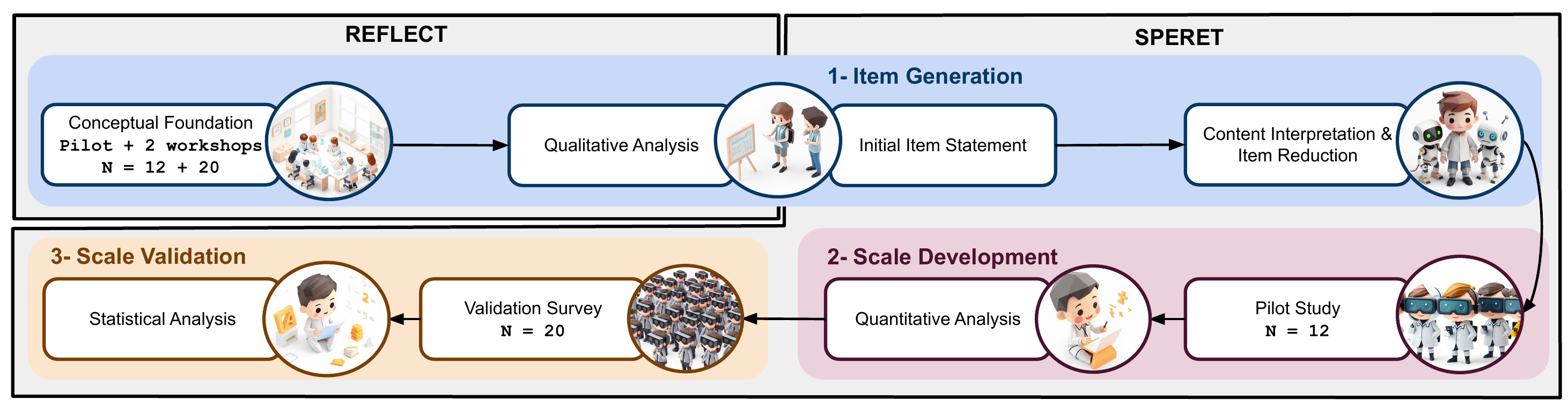}
    \caption{The development and linkage of our complementary tools REFLECT and SPERET. The thematic findings of the qualitative analysis are the basis for the item statements. $N$ denotes the number of participants. The icons are generated by deepai.org.} %
    \label{fig:method}
    \Description[Overview of our methodology with the REFLECT questionnaire, and the SPERET scale.]{The figure shows 8 steps in our approach each with a description and a little icon with avatars. The steps are contained in 2 boxes with the names of the tools REFLECT and SPERET, they also share three different background colours representing the threes stages of scale development (1- Item Generation, 2- Scale Development, 3- Scale Validation). The first step is Conceptual Foundation with a pilot + 2 workshops and N=12+20 participants, which is contained fully in the REFLECT box, and has the blue background colour for item generation. The second step is Qualitative Analysis, the icon for this step is on the border between the REFLECT and SPERET box, as the qualitative analysis forms the basis for the third step Initial Item Statement, which is the first step contained in the SPERET box. The remaining 4 steps are in the SPERET box. The fourth step is Content Interpretation \& Item reduction, which is the last step with blue background representing item generation. The fifth step is Pilot Study, and together with the sixth step Quantitative Analysis they share the red background representing Scale Development. The seventh step is Validation Survey with N=20 participants. The eighth and last step is Statistical Analysis. The last two steps share the yellow background representing the third stage Scale Validation.}
\end{figure}

To systematically understand, explain, and quantify privacy and ethical considerations within eye-tracking research, we developed two tools: the qualitative REFLECT questionnaire and the quantitative SPERET scale. 
These complementary tools foster and examine the reflexivity in eye-tracking research. We involved more than 70 eye tracking researchers in the tool development to ensure that the tools are anchored in eye-tracking researchers' reflexivity. This is particularly important for the quantitative scale, as previous work has shown that there is a discrepancy between what existing privacy scales measure and how their items are interpreted by the general public \cite{colnago2022concern}. 
To overcome this limitation we formulated the items of the SPERET scale based on the open reflective answers we collected with the REFLECT questionnaire. This linkage and the overall approach is shown in \autoref{fig:method}

In this section, we describe the development of both tools in detail outlining the rationale, procedures, and analytical strategies employed.

\subsection{REFLECT questionnaire}
The goal of the REFLECT questionnaire was to conceptually explore the reflexivity of eye-tracking researchers.
We wanted to do so by engaging researchers in a reflective practice that would be mutually beneficial—both for their own professional awareness and for our study. To this end, we organised a workshop that brought together researchers to learn about ethical and privacy considerations, while also providing dedicated time and space to reflect on their individual projects and share these reflections as part of our research.

In the following, we first explain the development of the REFLECT questionnaire, which served as the foundation for the reflective practice, and whose results later formed the basis for the item development for the SPERET scale. We then outline the workshop format and participants, and finally present our analysis of the collected responses, which informed the development of the resulting items.

\subsubsection{Development of REFLECT questionnaire}
The REFLECT questionnaire was adapted from a set of social-aspect questions in the Handbuch integrierte Technikentwicklung \cite{Spindler2026}, translated from German to English (see \autoref{questions_pilotversion}). We piloted the initial version with \textbf{N=12} eye-tracking and HCI researchers.

Pilot feedback indicated that fundamental research projects were difficult to map onto questions originally targeting applied technology development. To improve relevance, we replaced “users” with “target groups” (including both end-users and researchers), and added a research ethics question about participants in research projects to clearly separate target groups from research participants. We also formulated a question on non-intended use to support the researchers in anticipating how their work could be applied and also misused.  

We further identified two gaps. First, the absence of items probing underlying ethical assumptions and values was addressed by adding sections on conceptions of humanity, and societal relevance. Second, to capture challenges in machine learning–based eye tracking, we included a question on training data quality to elicit reflection on the situatedness of the data basis in machine-learning models. %

The questionnaire was iteratively refined by two authors and reviewed by a political scientist to ensure clarity and comprehensiveness. The final version comprises ten thematic blocks with 3–10 reflective questions (for an overview see \autoref{tab:reflect}).

\begin{table}[h!] 
\centering \caption{Excerpt of REFLECT Questionnaire. The questionnaire is designed to elicit reflection on ethical and privacy-related aspects of eye-tracking projects based on \cite{Spindler2026}. The questionnaire covered 10 thematic blocks, their theme is given in the left column. For each block there were 3-10 questions to guide participants. They were free to decide which ones to answer. In the right column exemplary questions are given, for the full questions see \autoref{questions_REFLECT}.    %
} \label{tab:reflect} 
\begin{tabular}{p{5cm}p{9cm}} 
\toprule 
\textbf{Theme} & \textbf{Exemplary Questions} \\ 
\midrule 
Underlying concept of humanity. & What would you say characterises "the humans" in contrast to other beings? [...] What do you consider to be the most important social/ethical/legal values that need to be preserved/promoted/protected?  %
\\[0.1cm] 
\midrule 
The conceptions of (future) target groups that influence your project. & Which groups of people, characteristics, behaviours, and prior knowledge are assumed? [...] Who haven't you thought of yet?%
\\[0.1cm] 
\midrule 
Ideas about good human coexistence that influence your work. & Which ways of living and working are considered? Which human needs do you have in mind? What is deemed desirable? 
\\[0.1cm] 
\midrule 
Societal relevance of your project. & %
[...] Why is your project important for this society? [...]
Explain it as you would explain it to a child or a non-academic, non-technical friend/family member. 
\\[0.1cm] 
\midrule 
Quality of training data. & What data is used to train the system? Who is generating the data? [...] %
Who has access to the training data and how? 
\\[0.1cm] 
\midrule 
Research ethics aspects. & Who else is involved in the development of the system apart from your working group? %
[...] Can participation in your research cause harm? What share do the participants have in their system? 
\\[0.1cm] 
\midrule 
The generation and meaning of data. & In which way does your project generate data? Who is generating the data (humans? sensors?)? Who is interpreting them? 
Which data do you generate but don't need them? %
\\[0.1cm] 
\midrule 
The design of interaction between humans and technology. & How are spaces of action and power relations structured? How is responsibility distributed between humans and technology? How is information distributed and made accessible/understandable/interpretable for humans? Who has access to which information? [...] %
\\[0.1cm] 
\midrule 
Non-intended use. & How should a system that can be derived from your project be used? [...] %
Can your project results be used in a different way than you intended? %
\\[0.1cm] 
\midrule 
Allocation of responsibilities. & Can your system cause damage? [...] %
Who (e.g a defined person, an institute, ...) is responsible for correcting mistakes or claims for compensation? 
\\ \bottomrule 
\end{tabular} 
\end{table}

\subsubsection{Workshops}
We organised two interdisciplinary workshops that addressed privacy and ethics in eye tracking: one at an academic eye-tracking conference and another one at our local university. The workshops provided additional input sessions to the participants, and thus they gave the REFLECT questionnaire the suitable framing of being developmental self-assessment instrument to support critical engagement with their hidden assumptions.

Participants completed the REFLECT questionnaire individually and by hand. The workshops provided dedicated time and space for them to record their thoughts on ethical and privacy aspects of their work using pen and paper. We intentionally chose this analogue format over a digital survey to encourage a slower, more deliberate writing process and deeper reflection, free from digital distractions. The participants responded positively to this approach, and during both workshops, the room was characterised by a calm atmosphere of focused concentration.

\subsubsection{Workshop participants}
Participants were researchers attending one of two workshops held at an eye-tracking conference (Conference Workshop: N=12) and our local university (Local Workshop: N=13). %
Five of the conference participants opted out of completing the REFLECT questionnaire, for timely reasons.
They are excluded from the following analysis, resulting in \textbf{N=20} participants overall. Participation was voluntary, and responses were anonymised for analysis.
The participants held different positions (8 master students, 5 PhD students, 4 professors, 2 Postdocs, and 1 engineer), they were more men than women (13 men, 7 women). The majority held a degree in computer science (12), with others having educational backgrounds in psychology/cognitive science (3), linguistics (3) and engineering (2). We also asked for the participants' cultural and social roots. The majority of the participants indicated German (6) or European (3) roots, individual participants also reported Native American (1), Spanish American (1), US-Japanese (1), Indian/Hindu (1), Romanian (1), Pakistani (1) and Egyptian (1) roots. %

\subsubsection{Data Analysis}
We followed a rigorous analysis pipeline to examine the open-ended questionnaire responses. First, all responses were transcribed and anonymised to ensure data integrity and participant confidentiality. We then conducted a qualitative thematic analysis based on the set of methods of grounded theory \cite{corbinBasicsQualitativeResearch2025} and using a hybrid inductive–deductive approach \cite{farzandOutofDevicePrivacyUnveiled2024, hasan2023psychometric}. The high-level structure of the REFLECT questionnaire provided initial deductive codes (e.g., 'Values', 'Target Group Assumptions'), within which we inductively identified emergent themes and patterns from the response data. 

The analysis was an iterative process of coding, discussion, and refinement. %

\subsubsection{Thematic Findings}

Our analysis identified three central themes that characterise the ethical reflections of the eye-tracking research community.%
    \\
\textbf{Theme 1: The Primacy of Data Privacy} 
\\
    Privacy emerged as the most salient ethical concern among participants. This emphasis reflects the focus of previous considerations (see \autoref{sec_privacyrelatedwork}) as well as the context of the workshop centred on privacy and ethics in eye tracking in which the questionnaire was completed. %
    Some responses directly echoed this setting, as illustrated by \textbf{P2}: \textit{“Privacy is a key social/ethical/legal value needing protection \& promotion.”} Similarly, \textbf{P20} identified privacy as the field’s foremost ethical challenge: \textit{“Our most important ethical challenge is data privacy.”} 
    Other participants provided more elaborate justifications that connected privacy to dignity, wellbeing, and legal protection. \textbf{P3} wrote: \textit{“Humans have a complex social system, where the concept of ‘individual dignity’; it is an important determinant of their overall wellbeing.”} The same participant elaborated on this by linking privacy to autonomy and intentional self-presentation: \textit{“Humans should have the right to present themselves in a way that preserves their dignity: consciously \& intentionally. Unconscious self-expression should be private.”} 
    Participants also referred to technical safeguards. \textbf{P18} noted: \textit{“The security aspect should be preserved, e.g. at least pseudomisation of data.”} 
    We also found competing values in addressing the special sensitivity of gaze data in two particpants' answers.
    \textbf{P1}, who worked on a learning environment using eye-tracking to assess student performance, highlighted the privacy of the recordings: \textit{“Only the concerned student can review/rewatch his/her simulation.”} They also worried about teachers accessing unintended information: \textit{“Teachers can find out things they wouldn't in a ‘normal situation’.”} 
    Interestingly, \textbf{P9} viewed this sensitivity as a reason to limit data access for the users themselves: \textit{“Information about body reactions/instinctive reactions can be dangerous [...] So the user should not directly access the system output but benefit from experts using the eval[uation] result to improve the quality of output of other user facing systems.”}. 
    The primacy of data privacy translated into practical design decisions involving access control and data anonymisation. For example in response to our question "Who has access?", \textbf{P4} reported: \textit{“Just myself in pseudonymized form; we keep a list of pseudonymized codes + personal information w/ organization + technical measures to prevent non-authorized users to access the data.”} Likewise, \textbf{P5} emphasised secure storage: \textit{“Only develops, secure physical storage.”} 
    \\
\textbf{Theme 2: The Assumption of Generalizability and the WEIRD Participant Burden}

    Many participants assumed that findings from their eye-tracking research generalise across human populations. \textbf{P13} made this assumption explicit: \textit{“We assume that all humans... are very very similar in their basic perception and cognition.”} Similarly,
    \textbf{P8} noted: \textit{“We do assume that future participants behave similarly to our tested sample, i.e., that our results generalise, which they may not."}. We also identified generalising statements that rest on this assumption without stating it directly: \textbf{P19} wrote: \textit{“Society will benefit from the project because it might help us understand spoken language processing and general human behaviour better [...]”}. However, this participant’s project focused specifically on English language processing, yet they generalise from this linguistic or cultural context to "spoken language processing and general human behaviour".
    The assumption is necessary to underpin the perceived scientific validity of their work but is undermined by the well-documented WEIRD (Western, Educated, Industrialised, Rich, Democratic) sampling bias \cite{blignautEyetrackingDataQuality2014, angeleClosingEyetrackingGap2024}. Some participants demonstrated awareness of this issue. As \textbf{P8} noted: \textit{“Self-selection bias, findings may not apply.”} 
    For others, the narrowness of their sample was acknowledged but unexamined. \textbf{P12} admitted: \textit{“Mostly students volunteering because we know them.”} 
    This uncritical reliance on convenience sampling reinforces the WEIRD bias and limits the external validity of eye-tracking studies \cite{angeleClosingEyetrackingGap2024, kaufmannEthicalChallengesResearching2021}. %
    \\
\textbf{Theme 3: Misuse, Unintended Use, and Reasons Not to Consider It}
\\
    When prompted about unintended applications or misuse, participants envisioned several possible risks: \\
    \textit{\textbf{Monitoring and guiding attention:}} \textbf{P2} warned, \textit{“It could be misused to police children who are chronically disengaged.”} Similarly, \textbf{P13} foresaw surveillance uses: \textit{“It would also be developed to monitor interests of humans (e.g. What happens if I look at object X) and in turn help guide attention towards specific things.”} \textbf{P5} mentioned commercial misuse: \textit{“Interest detection for ads.”} \\
    \textit{\textbf{Deceptive interfaces:}} \textbf{P15} cautioned, \textit{“Companies can use [it] to represent their data in a way it only reflect good things. The data representation... will not be fair.”} \\
    \textit{\textbf{Revealing private information:}} \textbf{P6} reflected, \textit{“[It] could be used in a bad way if data is analysed against different metrics—a lot of information could be revealed about the users without their consent.”}
\\
    These reflections indicate that participants are capable of identifying potential harms when explicitly guided to do so. %
    However, some participants struggled to envision either misuse or practical applications of their research. \textbf{P11} wrote: \textit{“I cannot think of a deviated way of using our project results, but the odds exist.”} \textbf{P13} expressed a similar sentiment: \textit{“I don't see any points to use our project results in a different way. The project is about language and brains, it's related to cognitive science but not a ground-breaking project.”} Likewise, \textbf{P8} dismissed the relevance of the question of possible unintended consequences of the developed system : \textit{“There is no ‘system’.”}
    These responses suggest a perceived divide between foundational and applied researchers that was also reported in other fields \cite{LiuEthicsStatements}. Those who identify as basic researchers tend to frame their work as neutral, curiosity-driven, and ethically unproblematic. For example, \textbf{P8} explained: \textit{“We're doing basic research, not designed to be immediately ‘useful’.”} Similarly, \textbf{P19} wrote: \textit{“Society will benefit from the project because it might help us understand spoken language processing and general human behaviour better while not being an invasive experiment.”} 
    By contrast, applied researchers 
     more readily engaged with ethical and societal implications, recognising both benefits and risks: \textbf{P3:}\textit{“Data holders place the onus of the individual's wellbeing on the individual; [...] %
    users benefit from centralized privacy protections \& increased usability."} and \textbf{P6:}\textit{"There will be security breach \& this should be accounted for from the beginning."}

Together, these themes reveal an ethical landscape in which eye-tracking researchers are highly attuned to issues of privacy, less critically reflective about the limits of generalisability, and uneven in their engagement with the broader social consequences of their work. The findings contribute in two respects.
First, they offer empirical insights of how eye-tracking researchers perceive privacy and ethics aspects in their work. Second, they form the basis for the initial item statements of our psychometric scale SPERET. 
This approach anchors the items firmly in eye-tracking research because the thematic findings are taken directly from the participants of the workshops.

\subsection{SPERET scale}
To facilitate the practical use of the thematic findings from our REFLECT questionnaire, we developed a psychometric scale that translates its key themes into measurable components, eliminating the need for complex qualitative analyses. 

For the scale development we adopted a well-established three-stage methodological framework inspired by Farzand et al. \cite{farzandOutofDevicePrivacyUnveiled2024} and Hasan et al. \cite{hasan2023psychometric}, the stage of each step is indicated by the background colour in Figure \ref{fig:method}.

\subsubsection{Stage 1: Item Generation}
Based on the thematic findings in the qualitative REFLECT data, we created initial item statements. For each thematic finding, we included multiple statements that were based on the participants' answers we collected. When there were multiple nuances in one thematic finding, we included more items to capture those. These resulted in 
initial item statements with \textbf{$46$ }items. %
In the next step, we iterated these initial statements to refine the item pool and evaluate content validity with three evaluators%
: two AI evaluators (GPT-5 \cite{openai_chatgpt_2025} and DeepSeek-R1 \cite{deepseekR1_2025}) %
and one eye-tracking and human–computer interaction (HCI) postdoctoral researcher with high English proficiency to review all items. The reviewers assessed four key aspects of content validity (as recommended by previous work \cite{farzandOutofDevicePrivacyUnveiled2024, worthingtonScaleDevelopmentResearch2006} %
: (1) identification of duplicate or redundantly worded items, (2) verification of item relevance to the construct definition, (3) evaluation of potential subjective interpretation, and (4) inspection of linguistic clarity and accuracy.

Of the $46$ initial items, $4$ were identified as duplicates and $2$ as irrelevant to the construct. The remaining items were reviewed and revised, where necessary, to improve linguistic precision and reduce ambiguity, resulting in a final pool of $40$ items (see \autoref{tab:40} for the full list).

\subsubsection{Stage 2: Scale Development}
The preliminary set of statements was tested in a pilot study with a sample of \textbf{(N = 12)} eye-tracking researchers to evaluate the clarity, interpretability, and psychometric potential of the items before large-scale data collection. This pilot phase was a crucial step in the iterative development of the questionnaire, aiming to ensure that the items were both conceptually sound and practically comprehensible to members of the target group. 

Specifically, the pilot testing focused on two key aspects:
(1) Comprehensibility of statements – assessing whether participants understood each item as intended, including the clarity of wording, alignment with the construct, and the adequacy of the response format; and
(2) Response variability – examining whether the items elicited a sufficient range of responses across the Likert scale \cite{Likert1932} to capture meaningful individual differences and to avoid ceiling or floor effects.

Overall, participants reported that the statements were clear and easy to interpret, and no major ambiguities or inconsistencies were identified. Based on participant feedback and preliminary response analyses, minor wording adjustments were made to improve precision and readability. This serves as an additional validation of the items as the participants were eye-tracking researchers, too.

To evaluate the reliability of the developed scale, we first examined the item–total correlations, which quantify the degree to which each individual item correlates with the sum of all other items in the scale (c.f. Table \ref{tab:40}). High item–total correlations indicate that an item is consistent with the overall construct being measured, whereas low correlations suggest that the item contributes little to the scale’s internal consistency and may not adequately reflect the underlying construct. Following standard psychometric guidelines \cite{boateng2018best}%
, we removed $14$ items whose item–total correlation coefficients fell below $0.30$, as such values are generally considered indicative of weak associations with the overall construct. This step ensured that the remaining items coherently represented the intended theoretical domain.

Subsequently, we assessed internal consistency reliability using Cronbach’s alpha \cite{Cronbach_1951}%
, a widely accepted metric of the extent to which items in a scale are interrelated and measure the same underlying construct. The analysis revealed a Cronbach’s alpha of $0.874$ for all items prior to item removal, indicating good overall reliability. After removing the low-correlation items, Cronbach’s alpha increased to $0.936$, demonstrating very high internal consistency. These results indicate that the refined set of items reliably measures a single, coherent construct—in this study, researchers’ privacy and ethics reflexivity.

Overall, this two-step reliability evaluation—combining item–total correlations and Cronbach’s alpha—provides empirical support that the scale is internally consistent and suitable for subsequent use in research assessing ethical and privacy reflexivity among researchers.

\subsubsection{Stage 3: Scale Validation}

To establish the reliability and validity of the developed scale, we conducted a validation study with an additional sample of 20 eye-tracking researchers \textbf{(N = 20)}. Recruiting participants with such specific expertise posed a substantial challenge, resulting in a small but specialized sample. Participants were 9 men, 8 women, and 3 who preferred not to say; aged 18–60 years (2 aged 18–30, 16 aged 31–40, 1 aged 41–50, 1 aged 51–60). Educational backgrounds included BSc (1), MSc (5), PhD (6), postdoc (6), and professor (2). Most worked in academia (16) or in eye-tracking companies (2). Participants were based in Africa (2), Asia (1), North America (2), and Europe (15). Research orientations were foundational (3), applied (5), and both (8).

Given this limited sample size, traditional confirmatory factor analysis (CFA) or structural equation modeling (SEM)–based fit indices (e.g., RMSEA, CFI, TLI, or $\chi^2$) commonly employed in prior psychometric studies would be statistically unreliable and unstable, as these indices depend on large-sample asymptotic assumptions. Therefore, instead of relying on global model fit indices, we focused on assessing internal consistency reliability and computed Cronbach's alpha coefficient for the validation study sample. The value of $\alpha$ = $0.772$ indicates satisfactory internal consistency of the scale, suggesting that the items coherently measure the intended construct within this expert population, as shown in Table \ref{tab:final}.

\begin{table}[h!]
\centering
\small
\caption{The final scale. To administer the scale, the items should be presented on a 7-point Likert scale, ranging from strongly disagree to strongly agree. The items are clustered into subscales matching the thematic findings privacy, sampling bias, and misuse, which can be combined or used individually. All statements within a subscale are required to be answered, and none involve reverse scoring. To minimise potential order effects, the presentation of items should be randomised. The overall scale score can then be obtained by computing the mean of all item responses, representing an aggregate measure of the respondent’s level of privacy and ethics reflexivity.}
\label{tab:final}
\begin{tabular}{p{15cm}}
\toprule
\textbf{Item} \\
\midrule
\textbf{Data Privacy:} \\
It is important for me that gaze data is treated as highly sensitive personal information.\\

I worry that my research data could be accessed or misused beyond its original purpose.\\

I believe that anonymisation alone is not sufficient to protect participants’ privacy.\\

I believe gaze data deserves stronger protections than other types of biometric data.\\

I am concerned about the possibility of re-identification even after anonymisation.\\

I feel responsible for ensuring that my participants’ data cannot be misused.\\

I strongly value protecting participants' autonomy and informed consent.\\

\textbf{WEIRD Participant Burden:} \\

I sometimes feel ethically conflicted about continuing research despite sampling biases.\\

I believe my research is weakened by relying mostly on student participants.\\

I worry that participant homogeneity leads to biased or unfair technologies.\\

I would prefer to work with more diverse samples even if it complicates my research.\\

I worry that scientific findings based on eye tracking depend on subjective decisions by the involved researchers.\\

I believe fairness requires ensuring that technologies work across diverse populations. \\

\textbf{Misuse Fears:} \\
I worry that eye-tracking research could be misused for surveillance purposes.\\

I believe that my work could unintentionally be used to manipulate users in advertising or interfaces.\\

I feel anxious about the possibility of my research enabling harmful downstream applications. \\

I believe there is a real risk that my research could be misused in harmful ways.\\

I fear that my findings could be exploited for manipulative commercial purposes. \\

I worry about governments or institutions using eye-tracking for surveillance. \\

I feel that preventing misuse of my work is partly my responsibility.\\

I sometimes find it difficult to connect my foundational research to broader ethical responsibilities.\\

I feel that even basic research should anticipate potential misuses in the future.\\

I feel that foundational research should still consider future societal consequences.\\

I sometimes avoid thinking about misuse because my research is not application-oriented. \\

I trust that other actors (e.g., companies, policymakers) will handle ethical issues downstream.\\

I believe research should prioritise doing good for society, not just advancing knowledge. \\

\bottomrule
\end{tabular}
\end{table}

Our aggregated results on the validation set (Mean = $4.82$, Standard Deviation = $0.64$) indicate that participants generally exhibit a good level of privacy and ethics reflexivity (slightly above the scale midpoint of $4$), with limited variability across individuals. This suggests a shared awareness - although limited - of ethical considerations within the surveyed eye-tracking researchers. 
These findings provide preliminary support for the internal coherence of the scale: responses cluster around a common construct of ethical reflexivity. %

\label{sec:methodology}

\section{Discussion}

This discussion interprets our findings on ethical reflexivity in eye-tracking research, pointing to emerging patterns, variations in individual researcher practices, possible applications of our questionnaire and scale, and directions for future work.

\paragraph{\textbf{Qualitative Findings:}} Our REFLECT findings reveal that eye-tracking researchers demonstrate ethical awareness characterised by strong concern for user privacy, recognition of methodological limitations, and varying degrees of responsibility depending on research context. Privacy considerations have become integral to the community’s identity, yet researchers differ in how they conceptualize gaze data sensitivity—some viewing feedback as empowering, others as potentially distressing—highlighting that ethical decision-making involves carefully balancing competing values to identify appropriate technical and procedural safeguards.
Concerns about limited participant diversity reflect methodological reflexivity and a tension between striving for generalisability and valuing situated insights, with implications for fairness as eye-tracking technologies enter applied settings. Finally, while many researchers acknowledge potential misuse, others distance themselves from downstream consequences by emphasizing the basic nature of their work, revealing a divide between theoretical and applied domains in ethical accountability.

\paragraph{\textbf{Quantitative Findings:}} These findings were highlighted quantitatively by our SPERET scale. Our analysis indicates that participants exhibit a moderately high level of privacy and ethics reflexivity, with relatively low inter-individual variability. This pattern suggests a broadly shared—though not exhaustive—awareness of ethical considerations among the surveyed eye-tracking researchers. The clustering of responses around a common range provides preliminary evidence for the internal consistency of the scale, supporting its utility as a measure of ethical reflexivity.

\paragraph{\textbf{Acknowledging Proactive Mitigation in Eye-Tracking Research}}
A critical nuance that emerged from our analysis is that, while the eye-tracking research community as a whole exhibits a moderately high level of privacy and ethics reflexivity, this tendency is not uniform. 
Two researchers (P3 and P6) showed notably higher anticipatory reflexivity, treating ethical risks as concrete technical and governance challenges. Their work incorporates proactive mitigation strategies such as gaze data encryption, accountable data-use frameworks, and privacy-aware data collection methods. This aligns with emerging literature on formalising privacy-preserving eye-tracking techniques (see Section~\ref{sec_privacyrelatedwork}).
This pattern suggests that targeted expertise and focus on ethical considerations can significantly elevate reflexivity, bridging gaps observed in the wider research community.

\begin{figure}
    \centering
    \includegraphics[width=0.8\linewidth]{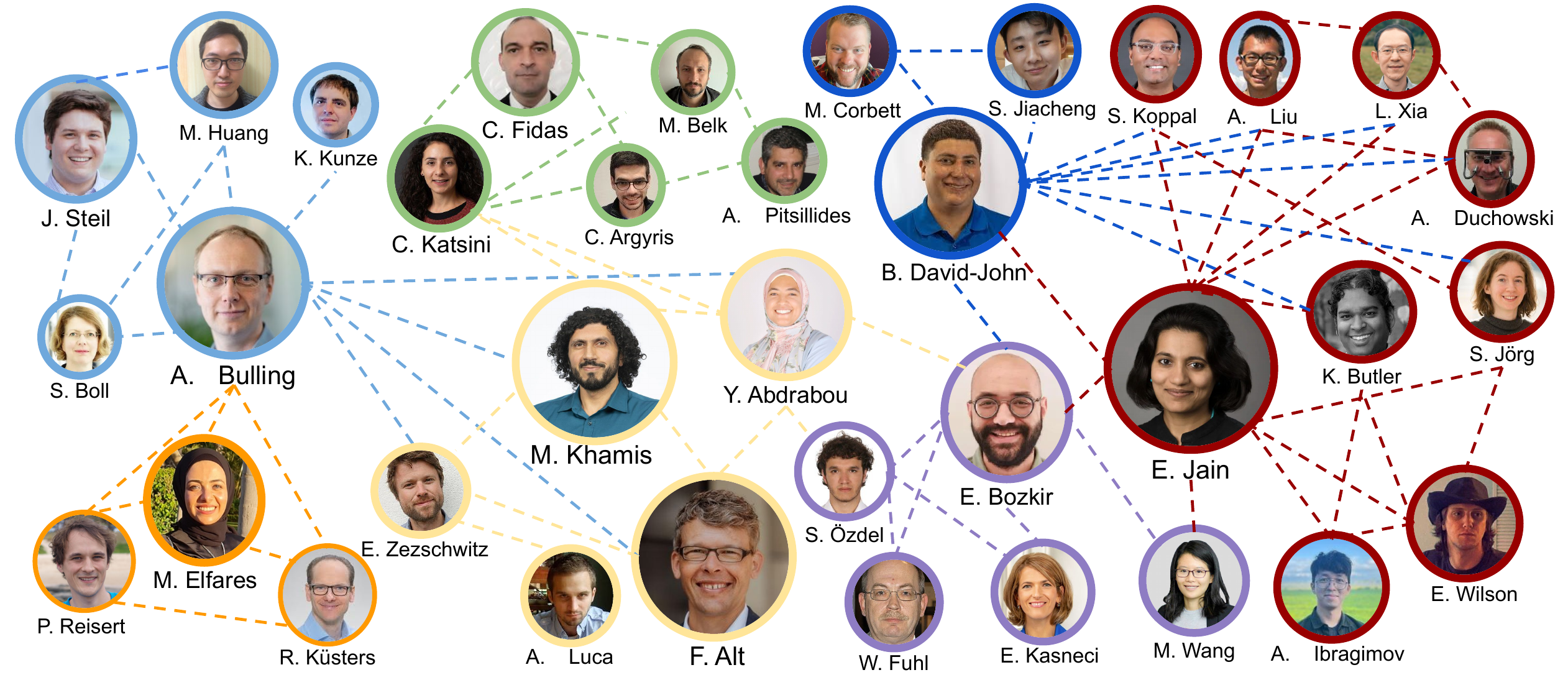}
    \caption{A recognition and documentation of researchers who have made significant contributions to the field of privacy-preserving eye tracking. Although they currently represent a minority within the broader community, our objective is to acknowledge their efforts and encourage continued advancement in this important area. The infographic is generated based on dimensions.ai's content.}
    \Description[Overview of authors who address privacy and ethics aspects in their work.]{The figure shows a network of authors represented with their name and portrait. It shows ca. 30 researchers. The most prominent ones are from left to right A. Bulling, M. Khamis, F. Alt, B. David-John, E. Jain}
    \label{fig:recognition}
\end{figure}

Therefore, we would like to acknowledge and document researchers who have made substantive contributions to the advancement of privacy-preserving approaches in eye-tracking research in Figure \ref{fig:recognition}. By recognising their efforts, we aim not only to highlight exemplary contributions but also to foster broader awareness and encourage sustained progress toward more privacy-conscious eye-tracking technologies and methodologies.

\subsection{Using Our Tools}
The REFLECT questionnaire and the SPERET scale offer actionable insights into the ethical and privacy reflexivity of eye-tracking researchers, providing both individual and community-level benefits. 

\paragraph{\textbf{At the individual level}} Completing the questionnaire and scale encourages researchers to identify and critically examine their own assumptions, biases, and ethical blind spots that may unconsciously influence study design, data collection, and interpretation. \textbf{P32} who participated in the scale validation study mentioned: '\textit{I do not know whether my collected gaze data can identify the participants, whether this is possible, and if there are tools that can do this.}'. This reflexive awareness can lead to immediate changes in research practices, such as adopting stronger privacy education and safeguards, reconsidering sample diversity, or integrating anticipatory ethical checks into project planning. For researchers, the value lies in strengthening the rigour, transparency, and societal accountability of their work while fostering professional development in ethical reasoning.

\paragraph{\textbf{At the community level}} Aggregated responses from REFLECT and SPERET can reveal shared patterns, gaps, and divergences in ethical reflexivity across the field. This information enables the design of targeted interventions—such as ethics workshops, collaborative reflection sessions, or updated institutional guidelines—that address systemic blind spots rather than isolated individual behaviors\footnote{During the workshops, we asked participants to rate how much initiatives like the privacy and ethics statements and our workshop affect their work. We noticed that the workshop had a slightly bigger effect than writing the statement. However, the design of such initiatives and their nuanced effect on reflexivity remains an open problem for future work.}. By establishing norms of reflexive practice, the REFLECT questionnaire and SPERET scale help cultivate a research culture where ethical considerations are integrated into methodological standards, peer review, and technology deployment.

\paragraph{\textbf{Our Questionnaire in Practice:}}
The REFLECT questionnaire can act as an ethical impulse in eye-tracking projects \cite{spindlerSituatedMultimodalityIntegrating2025}. 
We recommend that project members, particularly those from diverse disciplinary backgrounds, complete the questionnaire together in an analogue format and subsequently discuss and compare their responses. This collective, interdisciplinary reflection process helps to make explicit the often-implicit assumptions, values, and perspectives that shape a project’s direction. By surfacing these underlying viewpoints early, teams can prevent difficult or irresolvable negotiations later in the design process and reduce the risk of unintentionally embedding these assumptions into the resulting technology. Integrating insights from multiple disciplines, e.g. eye-tracking researchers and ethicists, further strengthens ethical reflexivity, ensuring that diverse viewpoints inform both design decisions and research practices.

\paragraph{\textbf{Our Scale in Practice:}}
Our scale can help in quantitatively answering questions such as '\textit{To what extent are eye-tracking researchers aware of the ethical and privacy implications of their work?}', '\textit{How well do researchers understand the potential risks associated with collecting, storing, and sharing gaze data?}' or
'\textit{To what degree do researchers perceive privacy as a core research concern rather than a peripheral consideration?}'. 
The presented results from item validation suggests that the surveyed researchers are aware of the privacy and ethics issues specific to eye tracking. This supports the findings from a large-scale survey that researchers consider ethical issues in their work \cite{berghaeuserResearchersPracticePerception2025}.
In future research, our scale could be applied to examine whether policy measures—such as the requirement to include societal impact statements—affect how researchers perceive ethical and privacy aspects in their work. Previous large-scale surveys have shown that institutional support enhances researchers’ engagement with research ethics \cite{berghaeuserResearchersPracticePerception2025}. Our scale could therefore be employed to investigate the specific impact of institutional support within the field of eye-tracking research. Additionally, it could serve to compare levels of ethical and privacy awareness across different subfields, for instance between researchers developing privacy-preserving techniques and the broader community of eye-tracking researchers.
\\

\subsection{Limitations and Future Work}
This study is constrained by a relatively small sample size given its specialized nature and potential self-selection bias, as researchers who contributed may already possess heightened ethical engagement relative to the broader eye-tracking community. Consequently, individuals less inclined toward ethical reflection may be under-represented, limiting the generalizability of our findings. Additionally, participants’ ethical assumptions and informal reasoning may have been influenced by contextual factors such as cultural background and educational experience. %
The sample size was insufficient to systematically examine these influences. Future research should address these limitations by engaging a larger and more diverse cohort of eye-tracking researchers, for example through qualitative analyses of societal impact statements or other reflexivity-focused interventions.

Additionally, future work should focus on expanding structured initiatives aimed at enhancing privacy and ethics awareness within the eye-tracking research community. Regular workshops, seminars, and training sessions led by experts from diverse domains—including ethics, data protection, human–computer interaction, and social sciences—can help researchers develop a more comprehensive understanding of the ethical implications of their work. Such interdisciplinary engagement would not only deepen individual reflexivity but also foster a shared vocabulary and culture of ethical responsibility across research teams and institutions. Integrating these sessions into existing research frameworks, conferences, and graduate education could further normalize ethical reflection as a core component of scientific practice, ensuring that privacy and ethical considerations are proactively addressed in eye-tracking research.%

Lastly, future work should explore whether the tools could be adapted to other research fields. Both tools strike a balance between specificity and generality: while they are rooted in the experiences of eye-tracking researchers, their wording and structure allow for potential adaptation to other research domains. This ensures that REFLECT and SPERET are both meaningful within the field and promising as models for broader applications. %

\subsection{Societal Impact Statement}

A potential risk associated with the use of our tools is the phenomenon of ethics washing (i.e. the performative adoption of ethical principles, intended to create the appearance of responsibility without enacting meaningful ethical practices or accountability), whereby participants may provide inflated or socially desirable responses to obtain higher reflexivity scores, particularly if such measures are used as evaluative criteria—for instance, in funding applications, institutional assessments, or admission processes. In such contexts, the instruments risk shifting from tools that foster genuine ethical reflection to performative mechanisms that reward compliance rather than critical engagement. This distortion undermines the instruments’ intended purpose of facilitating self-awareness and authentic reflexivity. To mitigate this risk, the questionnaires should be framed as developmental rather than evaluative tools, emphasizing their role in guiding discussion, learning, and ethical growth rather than measuring moral virtue or determining eligibility. Encouraging confidential, non-punitive use and embedding them in facilitated interdisciplinary reflection sessions can further ensure that the focus remains on cultivating ethical awareness rather than signalling it.

We finally note that, all data were collected voluntarily and handled anonymously to respect participant privacy and data sovereignty.
\label{sec:discussion}

\section{Conclusion}
In conclusion, the path to responsible eye-tracking technology begins with understanding the mindsets of its creators. Our study provides the first empirical map of this landscape within the academic eye-tracking community, through both qualitative (REFLECT) and quantitative (SPERET) tools. We found researchers were concerned for privacy and self-aware of the limitations, but also divided by a gap in anticipatory ethics between basic and applied research. 
Collectively, our tools and analysis offer both an assessment of reflexivity within eye-tracking research and a framework to encourage practices that are more attentive to privacy and ethical considerations.

\label{sec:conclusion}

\begin{acks}
\textbf{Author Contributions :}
The study was conceptualised and designed by Authors 1, 2, 3, and 4. Authors 1 and 3 led the preparation of workshop materials, organisation, and facilitation of the workshops. Authors 1 and 2 conducted the qualitative analysis, with support from Author 3. Author 2 developed and validated the measurement scale. Authors 1 and 2 led the writing of the manuscript, with input and critical revision from all authors. Author 4 provided overall supervision, conceptual guidance, and feedback throughout the research process.
\end{acks}

\bibliographystyle{ACM-Reference-Format}
\bibliography{references}

\newpage
\appendix
\section{ETRA Call for Privacy and Ethics Statement}
\label{ETRA_EthicsStatement}
\begin{shadedquotation}
With burgeoning technological and social applications of eye-tracking research, we encourage authors to consider potentially harmful impacts of their work and to try to mitigate any current or future societal risks that might result from its publication. Authors are required to include a privacy and ethics statement related to the study conducted in the methodology section and encouraged to consider expounding on broader impacts of their work in the discussion, limitations, or implications sections relating to privacy, fairness, safety, human rights, data sovereignty, and future adoption or misuse in the context of a benefit/risk assessment.\\
\textit{- ETRA 2024, Privacy and Ethics Statement, David-John and Elfares}
\end{shadedquotation}
\section{REFLECT Questionnaire}
\subsection{Original Version}
The following questions were used in a pilot test with N=12 researchers. These questions were taken from the questionnaire on social aspects in \cite{Spindler2026} and translated to English. 
\label{questions_pilotversion}
\begin{enumerate}
    \item \textbf{The conceptions of users that influence the development process.} Who do you believe will use your system? Which groups of people, characteristics, behaviors, and prior knowledge are assumed? Which ones are not considered?
    \item \textbf{The involvement of users in the development process.} Do you involve users in your development process? If yes, how are the participants and thus the participation processes selected? What status do the participants have in the project? What status is attributed to the results of the participatory processes?
    \item \textbf{Ideas about good coexistence that influence your work.} Which ways of living and working are considered? Which human needs do you have in mind? What is deemed desirable?
    \item \textbf{Concepts of society that influence the development process.} Which values and norms are considered applicable? How do you define society? Who do you have in mind?
    \item \textbf{The generation and meaning of data.} In which way do you generate your data? Who is generating the data? Who is interpreting them? Which data do you need? How do you decide on data generation? Which data do you generate but don´t need them?
    \item \textbf{The design of interaction between humans and technology.} How are action spaces and power relations structured? How is responsibility distributed between humans and technology? How is information distributed? Who has access to which information? Who can change the system settings?
    \item \textbf{Distribution of accountability.} How do you prevent misinformation? Who is responsible for the security of the system? How can responsibility be distributed? Who can decide on which steps? Who is held responsible in the event of faults?
    \item \textbf{Acquirement or refusal of a technology by the users.} Is the technology accepted? Are there types of unintended usage? Where does the technology encounter rejection
    \item \textbf{Inequalities in access to technology.} What role do age, gender, physical and mental disabilities, socioeconomic status, etc., play?
    \item \textbf{Impacts of technology on society.} How does technology affect people's coexistence, work processes, economic practices, political decision-making, and the environment?
    \item \textbf{Interrelationships between social and technological developments.} What expectations and demands does society place on technology? How do, for example, neoliberalism and surveillance technologies influence each other?
    \item \textbf{Effects on the environment.} Which non-human actors do you have in mind? Which machines, animals, artifacts, plants, or other parts of the environment will be affected by your development? In which way will they be affected?
    \item \textbf{Other thoughts and aspects.}
\end{enumerate}

\subsection{Workshop Version}
The following questionnaire was used in the final workshops with N=20 researchers. The italic highlight indicates questions and thematic blocks that we added after the piloting and hence represent a modification beyond translation to the original questionnaire taken from \cite{Spindler2026}.
\label{questions_REFLECT}

\paragraph{Explanatory header:}
The questions presented below are designed to elicit reflection on ethical, social, and privacy-related aspects of eye tracking projects. The selection of questions is intentionally varied, with some delving deeply into the specifics of the projects. It is not expected that all questions will be answered in detail. You are welcome to answer individual questions in more detail, others in keywords. The sub-questions are intended to provide inspiration and guidance, but they are not mandatory. The space provided is not related to the scope of the response. Please feel free to use additional sheets if necessary.
Finally, it should be noted that this form of reflection is not intended to provide definitive “right” or “good” answers. Rather, it is an opportunity to reflect and think further.

\begin{enumerate}
    \item \textit{\textbf{Underlying concept of humanity.} What would you say characterises "the humans" in contrast to other beings? How would you describe the relationship between humankind and the world? What do you consider to be the most important social/ethical/legal values that need to be preserved/promoted/protected? }
    \item \textbf{The conceptions of \textit{(future) target groups} that influence your project.} Which groups of people, characteristics, behaviours, and prior knowledge are assumed? What characteristics do the target groups have? How do they benefit from your project? \textit{Why do they interact with it or its results? Who else is affected by your project? How does your project or its results affect these actors?} Who haven't you thought of yet? 
    \item \textbf{Ideas about good human coexistence that influence your work.} Which ways of living and working are considered? Which human needs do you have in mind? What is deemed desirable?
    \item \textit{\textbf{Societal relevance of your project.} How would you define the society you have in mind? Why is your project important for this society? Who will benefit from your project? How? What are the beneficial aspects? Explain it as you would explain it to a child or a non-academic, non-technical friend/family member.}
    \item \textit{\textbf{Quality of training data.} What data is used to train the system? Who is generating the data? How and by whom is this data selected and checked for validity? Who has access to the training data and how?}
    \item \textit{\textbf{Research ethics aspects.} Who else is involved in the development of the system apart from your working group? Who is distributing tasks? How are the credits distributed? Can participation in your research cause harm? What share do the participants have in their system? }
    \item \textbf{The generation and meaning of data.} In which way does your project generate data? Who is generating the data (humans? sensors?)? Who is interpreting them? Which data do you need? How do you decide on data generation? Which data do you generate but don´t need them? \textit{Who can track the data on which the system bases its analyses during use? How? Can data generation be adapted to different tasks of the system? How and by whom?}
    \item \textbf{The design of interaction between humans and technology.} How are spaces of action and power relations structured? How is responsibility distributed between humans and technology? \textit{How is information distributed and made accessible/understandable/interpretable for humans?} Who has access to which information? Who can change the system settings?
    \item \textit{\textbf{Non-intended use.} How should a system that can be derived from your project be used? For what purposes will it be used and by whom? Can your project results be used in a different way than you intended? For what purposes is the use of your project results or its components suitable? How easily can parts of your system be modified or decoupled? }
    \item \textbf{Allocation of responsibilities. } Can your system cause damage? What happens if your system makes the wrong decisions? Who (e.g a defined person, an institute, ...) is responsible for correcting mistakes or claims for compensation?
\end{enumerate}

\section{The Initial 40-item Scale} 

\begin{table}[h!]
\centering
\small
\caption{The initial 40-item scale. Items with item-total correlation < $0.30$ were excluded from the final scale.}
\label{tab:40}
\begin{tabular}{p{15cm}p{2cm}}
\toprule
\textbf{Item} & \textbf{Correlation} \\
\midrule

\textbf{Data Privacy:} & \\
It is important for me that gaze data is treated as highly sensitive personal information. & 0.645847\\

I worry that my research data could be accessed or misused beyond its original purpose. & 0.772189\\

I believe that anonymisation alone is not sufficient to protect participants’ privacy. & 0.456091\\

\sout{I feel uneasy if gaze data is stored on external servers.} & 0.256391 \\

I believe gaze data deserves stronger protections than other types of biometric data. & 0.531404\\

I am concerned about the possibility of re-identification even after anonymisation. & 0.509177\\

I feel responsible for ensuring that my participants’ data cannot be misused. & 0.572885\\

\sout{I assume users understand the risks of sharing their gaze data.} & 0.157476\\

\sout{I believe academic environments are safer than commercial ones for handling sensitive data.} & 0.115806\\

\sout{I assume that once collected, data can be fully controlled through technical measures.} & -0.141868\\

\sout{I believe eye movements can be reduced to objective, measurable data.} & 0.157633\\

\sout{I value transparency in how data is collected, stored, and used.} & 0.212520\\

I strongly value protecting participants' autonomy and informed consent. & 0.413103\\

\textbf{WEIRD Participant Burden:} & \\
\sout{I feel concerned that my reliance on university students as participants limits the fairness of my research.} & -0.171263\\

\sout{I believe that findings based only on WEIRD samples may not generalise to diverse populations.} & 0.161349\\

I sometimes feel ethically conflicted about continuing research despite sampling biases. & 0.498684\\

I believe my research is weakened by relying mostly on student participants. & 0.538970\\

\sout{I feel uncomfortable assuming that results from WEIRD samples apply universally.} & 0.039091\\

I worry that participant homogeneity leads to biased or unfair technologies.& 0.479989\\

I would prefer to work with more diverse samples even if it complicates my research.& 0.816123\\

I believe fairness requires ensuring that technologies work across diverse populations. & 0.593764\\

\textbf{Misuse Fears:} & \\
I worry that eye-tracking research could be misused for surveillance purposes. & 0.659859\\

I believe that my work could unintentionally be used to manipulate users in advertising or interfaces.& 0.345170\\

I feel anxious about the possibility of my research enabling harmful downstream applications. & 0.734599\\

I believe there is a real risk that my research could be misused in harmful ways. & 0.624304\\

I fear that my findings could be exploited for manipulative commercial purposes. & 0.674689\\

I worry about governments or institutions using eye-tracking for surveillance. & 0.543049\\

I feel that preventing misuse of my work is partly my responsibility. & 0.767232\\

I worry that scientific findings based on eye tracking depend on subjective decisions by the involved researchers. & 0.532045\\

I sometimes find it difficult to connect my foundational research to broader ethical responsibilities. & 0.455937\\

\sout{I believe that applied researchers bear more direct responsibility for ethical consequences than foundational researchers.} & 0.206808\\

I feel that even basic research should anticipate potential misuses in the future. & 0.680634\\

\sout{I find it difficult to link theoretical research to real-world impacts.} & 0.096232\\

\sout{I believe applied research naturally carries greater ethical risks than foundational research.} & 0.044122\\

I feel that foundational research should still consider future societal consequences. & 0.724888\\

I sometimes avoid thinking about misuse because my research is not application-oriented. & 0.681263\\

\sout{I see my role as producing knowledge, not policing its applications. }& -0.472597\\

I trust that other actors (e.g., companies, policymakers) will handle ethical issues downstream. & 0.317843\\

\sout{I believe technological solutions (like encryption) can adequately address ethical risks.} & -0.462521\\

I believe research should prioritise doing good for society, not just advancing knowledge. & 0.658982\\

\bottomrule
\end{tabular}
\end{table}
\label{sec:appendix}

\end{document}